\documentstyle[12pt]{article}
\def\lb{\label}
\def\be{\begin{equation}}
\def\ee{\end{equation}}
\def\ba{\begin{eqnarray}}
\def\ea{\end{eqnarray}}

\def\p{\partial_}
\def\e{{\rm e}}
\def\cE{{\cal E}}

\begin{document}
\title{
   \begin{flushright} \begin{small}
     LAPTH-848/01  \\ DTP-MSU/01-10 \\
  \end{small} \end{flushright}
\vspace{1cm}
%%%%  Title %%%%
{\bf Conformal mechanics on rotating Bertotti-Robinson
spacetime} }
\vspace{1cm}
%%%%%  Authors  %%%%
\author{
   {\bf G\'erard Cl\'ement$^{a}$}
\thanks{Email: gclement@lapp.in2p3.fr}
 and {\bf Dmitri Gal'tsov$^{a,b}$}
\thanks{Email: galtsov@grg.phys.msu.su} \\ \\
%%%%%  Address  %%%%
%\address
$^{a}$Laboratoire de  Physique Th\'eorique LAPTH (CNRS), \\
B.P.110, F-74941 Annecy-le-Vieux cedex, France\\
%\address
$^{b}$Department of Theoretical Physics,\\
     Moscow State University, 119899, Moscow, Russia,}

%%%%%  Date  %%%%
\date{23 May 2001}
\maketitle
\begin{abstract}
We investigate conformal mechanics associated with the rotating 
Bertotti-Robinson (RBR) geometry found recently as the
near-horizon limit of the extremal rotating
Einstein-Maxwell-dilaton-axion black holes. The solution breaks
the $SL(2,R)\times SO(3)$ symmetry of Bertotti-Robinson (BR)
spacetime to $SL(2,R)\times U(1)$ and breaks supersymmetry in the
sense of
$N=4, d=4$ supergravity as well. However, it shares with BR such
properties as confinement of timelike geodesics and
discreteness of the energy of test fields on the
geodesically complete manifold. Conformal mechanics governing
the radial geodesic motion coincides with that for a charged
particle in the BR background (a relativistic version of the De
Alfaro-Fubini-Furlan model), with the azimuthal momentum
playing the role of a charge. Similarly to the BR case, the
transition from  Poincar\'e to global coordinates leads to a
redefinition of the Hamiltonian making the energy spectrum
discrete. Although the metric does not split into a product
space even asymptotically, it still admits an
infinite-dimensional extension of $SL(2,R)$ as asymptotic
symmetry. The latter is shown to be given by the product of one
copy of the Virasoro algebra  and $U(1)$, the same being valid
for the extremal Kerr throat.

\end{abstract}

\section{Introduction}
In the chain of $AdS_n/CFT_{(n-1)}$ dualities \cite{Ma} the
lowest case $n=2$ attracted recently much attention. The
underlying conformal theory is one-dimensional quantum
mechanics, which is intimately connected with two-dimensional
CFT (dual to $AdS_3$), from which it inherits a Virasoro
extension.  While higher-$n$ dualities are mostly interesting
in what they teach us about (flat space) gauge theories, the
$n=2$ case is interesting in its gravitational aspects --
two-dimensional gravity, black hole entropy, etc. From the
`brane' viewpoint, the $AdS_2$ geometry is encountered in the
throat limit of an extremal Reissner-Nordstr\"om black hole
which is given by the direct product $AdS_2\times S^2$ known as
the Bertotti-Robinson (BR) spacetime. The former is a 1/2 BPS
state of $N=2$ four-dimensional supergravity \cite{GiHu82,To83},
whilst the throat exhibits the full $N=2$ SUSY. The $AdS_2$
sector is encountered also in several intersecting branes of
ten and eleven-dimensional supergravities \cite{BoPeSk98} and
$D0$ branes in any dimensions \cite{Yo99,CaCaCaMi01}.

The $SL(2,R)$ isometry group of $AdS_2$ is the conformal group
on a line. One way to uncover the conformal mechanics
associated with BR is to analyse radial motion of neutral or
charged particles down the throat. It is appropriate to
distinguish the $AdS_2^0$ patch which covers only the exterior
part of the throat from the full geodesically complete $AdS_2$
space, which we call here $AdS_2^+$. Geodesic motion on
$AdS_2^0$ was shown \cite{Cl98,Ka99} to be related to the
relativistic version of the De Alfaro-Fubini-Furlan (DFF)
conformal mechanics \cite{DFF}. The DFF hamiltonian has a
continuous positive semi-definite spectrum and no zero energy
ground state. As was shown in \cite{GiTo99}, passing to
$AdS_2^+$ is equivalent to adding a potential term making the
spectrum discrete with a well-defined ground state. The DFF model
may be regarded as a two-particle Calogero model, and it was
suggested by Gibbons and Townsend \cite{GiTo99} that the $N=4$
superextension of the large $n$-particle Calogero model may
describe quantum extreme RN black holes. The related
superconformal models were discussed in
\cite{MiSt99,Az99,AkKu99,Wy99}. An alternative way to reach an
infinite-dimensional extension of the conformal mechanics
consists in revealing a `hidden' Virasoro symmetry, this may be
achieved both non-geometrically
\cite{BeVa94,Ku99,Ku00,Ra01,CaCaKl01} and by exploring the
asymptotic symmetry of BR (see below). Still the problem of a
holographic description of microscopic degrees of freedom of
extreme RN black holes remains open, which is perhaps
related to the existence of the mass gap \cite{MaSt97}
excluding non-singular classical deformations of spacetime. An
alternative interpretation of internal degrees of freedom of
extremal RN black holes was suggested in
\cite{MaMiSt99,Mi98,Vo99} as `fragmentation' of the throat.

A natural arena to explore the $AdS_2/CFT_1$ correspondence is
two-dimensional gravity \cite{St99,CaMi99,NaNa00,CaMi002,
Ca99,CaMi004,CaCa00,CaCaKlMi00,CaCaVa00, CaVa00,CaCar01}. The
major physical question is related to the entropy of
two-dimensional black holes. The analysis starts with revealing
that the asymptotic symmetry of $AdS_2$ is given by the
Virasoro algebra acting on a line instead of the
two-dimensional boundary of $AdS_3$. In this latter case, as was
shown by Brown and Henneaux \cite{BrHe86}, the deformations of
spacetime preserving the structure of the asymptotic boundary
form a projective representation of the Virasoro symmetry with
a (classically calculable) central charge (for a more recent
derivation using path integrals see \cite{Te01}). The
definition of the corresponding algebra of charges for $AdS_2$
is more subtle since the boundary of a one-dimensional manifold
has zero dimensionality. A way out was suggested by Cadoni and
Mignemi \cite{CaMi99} by introducing some time average for the
charges. Using this procedure the central charges for $AdS_2$
were obtained in \cite{CaMi99} and \cite{NaNa00} (with somewhat
different parameterization for spacetime deformations)
differing by a factor of two (for a detailed discussion see
\cite{CaMi002}). The central charge is presumably related to
the entropy of two-dimensional black holes via Cardy's formula
(as was suggested earlier by Strominger for the BTZ black hole
\cite{St98}). The correct match is straightforward in the case
of the value found in \cite{NaNa00}, but requires accounting
for the entanglement entropy due to the second component of the
boundary of $AdS_2$ \cite{CaCaKlMi00} to obtain the value of
\cite{CaMi99}. For a discussion of this discrepancy see also
\cite{Ca99,CaCaVa00,CaVa00}. The relevance of the $2d$ gravity
central charge to the entropy of the extreme four-dimensional
RN black hole is problematic, though it may be responsible for
a {\it variation} of the entropy for a charged black hole away
from extremality \cite{MaSt97,NaNa00}.

One can hope to make further progress in understanding the
puzzles of $AdS_2/CFT_1$ by exploring other backgrounds with
$AdS_2$ sectors. It was suggested by Bardeen and Horowitz
\cite{BaHo99} (BH) that the near-horizon geometry of {\em
rotating} four-dimensional black holes (for an earlier
discussion see \cite{Za98}) can also provide an arena for
holography. The isometry group of the Kerr throat is
$SL(2,R)\times U(1)$, where  $U(1)$ is a remnant of the
spherical symmetry of BR. Although the geometry can not be
presented globally as a direct product of $AdS_2$ with
anything, it still preserves some typical AdS properties. The
main difference with higher-dimensional rotating brane/black
hole spaces \cite{MaSt97,CvLa99,Be00,BePa00,HoHuTa99} is that
for the four-dimensional Kerr spacetime the AdS sector does not
factor out even asymptotically. (However, a counterexample of a
five-dimensional $AdS_5$ Einstein-Maxwell black hole was
constructed recently \cite{CaKlSa01}). Still the Kerr throat
geometry turned out to be too complicated for analysis, so no
holographic dual  was found in \cite{BaHo99}. In particular, it
inherits from the full Kerr spacetime the superradiance
phenomenon whose interpretation in the holographic context is
obscure.

Here we make some progress in this direction using a geometry
which possesses the same isometry $SL(2,R)\times U(1)$ as the
BH solution but does not contain the angular factors causing
the above complications. Our solution is non-vacuum, it is
supported by a vector field and a pseudoscalar axion. This
geometry was previously found as a particular near horizon
limit of rotating dilaton-axion black holes \cite{ClGa01}. Here
we show that it is an exact solution of the four-dimensional
Einstein-Maxwell-dilaton-axion theory, which is a consistent
truncation of the $N=4, d=4$ supergravity or an effective
heterotic string theory. Then we investigate properties
underlying the structure of the dual conformal mechanics. We
show that confinement of geodesics is manifest exactly in the
same way as in the BR case. The same is true for the frequency
spectrum  of the test scalar field on the geodesically complete
manifold: the spectrum is discrete and there is no
superradiance. Analyzing the radial part of the wave equation
we find that for the $RBR^0$ coordinate patch the underlying
conformal mechanics is again the relativistic De
Alfaro-Fubini-Furlan model found in \cite{Cl98} for the BR case,
with the angular momentum playing the role of the electric
charge of \cite{Cl98}. So the spectrum is continuous with zero
energy excluded. On the contrary, the spectrum is discrete and
bounded from below if one deals with the full geodesically
complete manifold. We construct explicitly the spectrum
generated algebra related to the conformal symmetry.

The conformal boundary of the rotating BR is a
three-dimensional spacetime which, unlike BR, does not
factorise globally into the product of a line and a 2- sphere.
Moreover, the conformally rescaled boundary is now singular. We
show that in spite of this the asymptotic symmetry of RBR is
infinite dimensional and is given by the product of (one copy
of) the Virasoro algebra (without a central charge) with a
circle. The same holds for the near-horizon geometry of the
extremal Kerr-Newman black hole. The generators of the infinite
symmetry contain azimuthal derivatives, which reflect the
rotating nature of the spacetime.

The plan of the paper is as follows. In section 2 we review our
new solution as well as some related ones and discuss
isometries and various coordinate patches. In section 3 the
geodesics and the modes of a test scalar field are analysed on
the full RBR spacetime. We show that all timelike geodesics are
confined while the equatorial null ones can escape to infinity
but in an infinite time. So the RBR spacetime is geodesically
complete. The frequencies of the Klein-Gordon modes are
discrete and form an equidistant spectrum. In section 4 an
associated conformal mechanics is presented. Section 5 is
devoted to asymptotic symmetries of both BR and RBR spacetimes.
We conclude in 6 with some general remarks.
\setcounter{equation}{0}
\section{RBR solution to dilaton-axion gravity}
We start by recalling  the Bertotti-Robinson (BR) metric  which
is a solution to the four-dimensional Einstein-Maxwell theory
\be\lb{br} ds^2 =- r^2\,d\tau^2 + \frac{dr^2}{r^2} +
d\theta^2+\sin^2\theta\,d\varphi^2, \ee supported by the Maxwell
field \be A=qrd\tau+p\cos\theta d\varphi, \ee where electric
and magnetic charges are subject to the condition $q^2+p^2=1$
according to the choice of the unit radius of the hyperboloid.
This solution is known to represent the near-horizon geometry of
the extreme Reissner-Nordstrom (RN) black holes and is a fully
supersymmetric solution to $D=4,\;{\cal N}=2$ supergravity
\cite{GiHu82,To83}. Alternatively, (\ref{br}) arises in the
Kaluza-Klein reduction of a five-dimensional vacuum solution
with zero dilaton (i.e. with equal electric and magnetic
strengths in four dimensions) \cite{ClGa01}.

Geometrically this is a product space $AdS_2\times S^2$, whose
AdS component can be viewed as due to the effective
cosmological constant in two dimensions resulting from the
Maxwell field. The coordinates (\ref{br}) cover only a half of
$AdS_2$, but can be continued to the full manifold via an
appropriate coordinate transformation. Namely, the
one-parameter family of metrics \be\lb{brb} ds^2 =-
(x^2+b)\,dt^2 + \frac{dx^2}{x^2+b} +
d\theta^2+\sin^2\theta\,d\phi^2, \ee satisfies the
Einstein-Maxwell equations with the potential
$A=qxd\tau+p\cos\theta d\phi$ for any real $b$. For $b>0$ the
domain of $x$ (initially $r\geq 0$) can be extended to the full
real line thus covering the entire $AdS_2$ manifold (for $b\leq
0$ the radial coordinate is restricted to $x\geq \sqrt{-b}$).
To describe different coordinate patches on $AdS_2$ it is
enough to consider $b=0,\pm 1$, we will call the corresponding
metrics as $BR^0,\,BR^\pm$ (our notation for the last two is
opposite to the one used in \cite{CaMi99}). The $BR^+$ space is
geodesically complete, while $BR^0,\,BR^-$ have a Killing
horizon at $x=0,\, 1$ respectively.

Bardeen and Horowitz \cite{BaHo99} (BH) observed that an
effective cosmological constant in two dimensions may also
arise from the Kaluza-Klein two-form in dimensional reduction
of a {\em four-dimensional} vacuum metric, namely the extremal
rotating Kerr black hole. It has been shown that the
near-horizon geometry of the Kerr black hole contains indeed
the $AdS_2$ sector, though the direct product structure of the
four-dimensional spacetime is lost: \be\lb{BHz} ds^2 =
\frac12(1+\cos^2\theta)\left[-r^2\,d\tau^2 + \frac{dr^2}{r^2}
+  d\theta^2\right] + \frac{2\sin^2\theta}{1+\cos^2\theta}
(d\varphi+rd\tau)^2. \ee It can be checked this this metric is
an exact solution of the vacuum Einstein euqations possessing
an isometry group $SL(2,R)\times U(1)$. It approaches
$AdS_2\times S^2$ near the polar axis for small $r$, but
for larger $r$  the non-diagonal $\phi-\tau$ term becomes
manifest. Here the domain of $r$ is also a semi-axis, and there
is a Killing horizon at $r=0$. The Killing vector
$\partial_\tau$ becomes spacelike inside the ergosphere which
is given by \be \cos 2\theta < 4\sqrt{3} -7. \ee By coordinate
transformation one can  rewrite the solution in a form similar
to (\ref{brb}): \be\lb{BHzb} ds^2 =
\frac12(1+\cos^2\theta)\left[-(x^2+b)\,dt^2 +
\frac{dx^2}{x^2+b} +  d\theta^2\right] +
\frac{2\sin^2\theta}{1+\cos^2\theta} (d\phi+xdt)^2, \ee for
$b>0$ this patch covers the entire hyperboloid. Similarly to
the case of AdS, we will label the three distinct patches
$b=0,\,\pm 1$ of the BH solution as $BH^{0,\pm}$. It was shown
in \cite{BaHo99} that $BH^+$ partly preserves such typical AdS
features as confinement of timelike geodesics and discrete
energy spectrum of a minimally coupled test scalar field.
However, the presence of $\theta$-dependent factors violates
this simple picture: a class of geodesics can escape to
infinity, and the Klein-Gordon spectrum also contains a
continuous part with an associated superradiance phenomenon.
Note that the BH spacetime does not admit Killing spinors.

Recently we have found \cite{ClGa01} another spacetime which
which lies between (\ref{br}) and (\ref{BHz}): the metric is
non-diagonal but does not contain $\theta$-factors. The price
of this is its non-vacuum nature, the solution was obtained as
the limit $M\to \infty$, $a \to 0$ ($J = Ma$ fixed) of the
near-horizon limit of extremal rotating dilaton-axion black
holes \cite{GaKe94}. The action of this theory, which is a
truncated bosonic sector of $D=4,\;{\cal N}=4$ supergravity or
toroidally compactified heterotic string, describes the
gravity--coupled system of two scalar fields (dilaton $\Phi$
and (pseudoscalar) axion $\kappa$), and one Abelian vector
field $A_{\mu}$, : \be \lb{ac} S = \frac{1}{16\pi}\int
d^4x\sqrt{|g|}\left\{-R - 2\partial_\mu\Phi\partial^\mu\Phi -
\frac{1}{2} e^{4\Phi} {\partial_\mu}\kappa\partial^\mu\kappa
-e^{-2\Phi}F_{\mu\nu}F^{\mu\nu}-\kappa F_{\mu\nu}{\tilde
F}^{\mu\nu}\right\} , \ee where\footnote{Here
$E^{\mu\nu\lambda\tau} \equiv
|g|^{-1/2}\varepsilon^{\mu\nu\lambda\tau}$, with
$\varepsilon^{1234} = +1$, where $x^4 = t$ is the time
coordinate.} ${\tilde
F}^{\mu\nu}=\frac{1}{2}E^{\mu\nu\lambda\tau}F_{\lambda\tau},\;
F=dA\;$. Our new metric looks like a `pure' rotating BR:
\be\label{bremda} ds^2=-r^2 d\tau^2+\frac{dr^2}{r^2}
+d\theta^2+\sin^2\theta (d\phi+r d\tau)^2, \ee which coincides
with (\ref{BHz}) near the polar axis.
This geometry  is supported by a non-zero Maxwell field which is
the radial magnetic field distorted by the rotation induced
Faraday effect \be\lb{a} A=A_\mu
dx^\mu=-\frac{\cos\theta}{\sqrt{2}}(d\phi+r d\tau), \ee
 and
by the linear axion \be\lb{kappa} \kappa  = \cos\theta, \ee the
dilaton field being identically constant (set to zero). The
normalization corresponds to the chosen unit radius of $AdS_2$.

First we want to demonstrate  that this limiting solution is an
exact solution of the field equations following from the action
(\ref{ac}). The non-zero components of the Einstein tensor
$G_{\mu\nu}=R_{\mu\nu}-g_{\mu\nu} R/2$  in the orthonormal
frame (denoted by barred indices) read \be G_{{\bar \tau}{\bar
\tau}}=-G_{{\bar r}{\bar r}} =G_{{\bar \theta}{\bar
\theta}}=\frac14 (3+\cos^2\theta),\quad G_{{\bar \phi}{\bar
\phi}}=\frac14 (3\cos^2\theta+1), \ee while the components of
the Maxwell stress-tensor (multiplied by $8\pi$) are \be
T_{{\bar \tau}{\bar \tau}}=-T_{{\bar r}{\bar r}} =T_{{\bar
\theta}{\bar \theta}}=T_{{\bar \phi}{\bar \phi}} =\frac14
(\cos^2\theta+1), \ee the mismatch being accounted by the axion
stress-tensor \be T_{{\bar \tau}{\bar \tau}}=-T_{{\bar r}{\bar
r}} =T_{{\bar \theta}{\bar \theta}}=-T_{{\bar \phi}{\bar \phi}}
=\frac14 \sin^2\theta. \ee

It is simple to check the validity of the Maxwell equations for
the EMDA theory
 \be
\nabla_\nu(F^{\mu\nu}+\kappa{\tilde F}^{\mu\nu})=0, \ee and the
axion equation \be \Box\kappa=2 F_{\mu\nu}{\tilde F}^{\mu\nu},
\ee (these are written for a vanishing dilaton field), as well
as the compensation of the two right-hand side terms in the
dilaton equation \be
2\Box\phi=-F_{\mu\nu}F^{\mu\nu}+(\nabla\kappa)^2. \ee

Alternatively, as was shown in \cite{ClGa01}, the metric
(\ref{bremda}) can be regarded as originating from
five-dimensional gravity coupled to dilaton and axion, or from
six-dimensional vacuum gravity. More recently, a similar
solution was obtained in the five-dimensional Einstein-Maxwell
theory with cosmological constant \cite{CaKlSa01}.

Both (\ref{BHz}) and our solution break the $SO(3)$ symmetry of
(\ref{br}). The remaining isometry group $SL(2,R) \times U(1)$
is generated by the four Killing vectors \ba\lb{kils}
L_0 & = & \tau\partial_\tau- r\partial_r  \,, \nonumber \\
L_1 & = & \partial_\tau\,,\nonumber \\
L_{-1} & = &
 (r^{-2} + \tau^{2})\partial_\tau -2r\tau\partial_r
-2r^{-1}\partial_{\phi}\,, \\
L_\phi & = & \partial_{\phi}\,. \nonumber \ea The first three
of these generate the $sl(2,R)$ algebra \be
\left[L_n,L_m\right]=(n-m)L_{n+m}\,, \quad m,n=0, \pm1. \ee
which is an isometry of $AdS_2$ space (and a conformal symmetry
of the line). However the four-dimensional spacetime is not a
direct product of $AdS_2$ with another two-dimensional space,
as reflected in the presence of the term $\partial_\phi$ in
$L_{-1}$.

An attractive feature shared by both metrics (\ref{bremda}) and
(\ref{BHz})
 is that the conformal boundary at infinity is a timelike
surface. Contrary to the BR case, however, the conformal
boundary is now a singular curved space-time. Writing the
metric as $ds^2=r^{2} ds_c^2$, one finds for (\ref{bremda}): \be
ds_c^2=-d\tau^2
+d\rho^2+\rho^2\left[d\theta^2+\sin^2\theta\left(d\phi+\frac{d\tau}{\rho}\right)^2\right],
\ee where $\rho=1/r$. The scalar curvature of this spacetime \be
R_c=\frac{3\sin^2\theta}{2\rho^2} \ee diverges on the boundary
$\rho\to 0$. Nevertheless, as we will see below, many features
of the BR spacetime which are relevant for $AdS/CFT$
correspondence still hold in our case.

The variables $\tau, r$ are Poincar\'e coordinates on $AdS_2$
which cover only half of the AdS hyperboloid in
three-dimensional spacetime. Global coordinates are introduced
via the change of coordinates \be r=\sqrt{1+x^2}\cos t +x,\quad
r\tau=\sqrt{1+x^2}\sin t, \ee transforming \be -r^2\,d\tau^2 +
\frac{dr^2}{r^2}= -(1+x^2)d t^2 + \frac{dx^2}{1+x^2}. \ee In
addition, the angular coordinate transformation \be
\phi=\varphi+ \ln\left|\frac{\cos t+x\sin t }
{1+\sqrt{1+x^2}\sin t}\right| \ee preserves the angular term
\be d\phi+rd\tau=d\varphi+xd t. \ee Applying this to our
initial $RBR^0$ solution (\ref{bremda}) we obtain the
geodesically complete metric $RBR^+$ \be\lb{nhz} ds^2 =
-(1+x^2)\,dt^2 + \frac{dx^2}{1+x^2} + d\theta^2 +\sin^2\theta
(d\varphi+xdt)^2, \ee or, in a more symmetric form with
$y=\cos\theta$, \be\lb{nhxy} ds^2 = -(1+x^2)\,dt^2 +
\frac{dx^2}{1+x^2} + \frac{dy^2}{1-y^2}
+(1-y^2)(d\varphi+xdt)^2 \ee This metric cover the full $AdS_2$
hyperboloid. Note that its boundary has two disconnected
components $x\to \pm\infty$, as in the $AdS_2$ case (contrary
to higher $n$ AdS).

Another useful coordinate transformation \be
r=\sqrt{x^2-1}\cosh t +x,\quad r\tau=\sqrt{x^2-1}\sinh t,\quad
\phi=\varphi + \ln\left|\frac{\cosh t+x\sinh t }
{1+\sqrt{x^2-1}\sinh t}\right| \ee takes the interval into the
$RBR^-$ form \be\lb{nhxy-} ds^2 =- (x^2-1)\,dt^2 +
\frac{dx^2}{x^2-1} + \frac{dy^2}{1-y^2}
+(1-y^2)(d\varphi+xdt)^2, \ee with $x\geq 1$ and a Killing
horizon at $x=1$.

These three coordinate patches, together with their rescalings,
may be combined in the following solution valid for any real
$b$: \ba\lb{nhxyb} ds^2 & = &- (x^2+b)\,dt^2 +
\frac{dx^2}{x^2+b} + \frac{dy^2}{1-y^2}
+(1-y^2)(d\varphi+xdt)^2, \nonumber \\
A & = & -\frac{y}{\sqrt{2}}\,(d\varphi + x\,dt)\,, \qquad \phi
= 0\,, \qquad \kappa = -y\, \ea where we have taken care that
the coordinate transformation $(r,\theta,\varphi,t) \to
(x,y,\varphi,t)$ reverses orientations ($\partial
y/\partial\theta = -\sin\theta < 0$) so that the sign of the
pseudoscalar $\kappa$ must be reversed when going from
(\ref{kappa}) to (\ref{nhxyb}).

Note that for $RBR^- \,(b=-1)$ one can make the complex
transformation \be t \leftrightarrow i\varphi\,, \quad x
\leftrightarrow -y\,, \quad ds^2 \to -ds^2\,, \quad A \to
-iA\,, \ee leading to the Bertotti-Robinson-NUT  solution:
\ba\lb{brnut} ds_4^2 & = & -(x^2 - 1)\,(dt + y\,d\varphi)^2 +
\frac{dx^2}{x^2 - 1}
+ \frac{dy^2}{1 - y^2} + (1 - y^2)\,d\varphi^2\,, \nonumber \\
A & = & \frac{x}{\sqrt{2}}\,(dt + y\,d\varphi)\,, \qquad \phi =
0\,, \qquad \kappa = x\,. \ea This solution breaks the
$SL(2,R)$ symmetry of $AdS_2\times S^2$, but preserves the
spherical symmetry, the isometry group being $U(1)\times SO(3)$.

All the solutions obtained from the extremal rotating EMDA black
holes \cite{GaKe94} are non supersymmetric in the context of $N=4, d=4$
supergravity \cite{ClGa01}. This is perhaps not surprising
since these black holes are not BPS states. Non-rotating
(and NUT-less) BPS EMDA black holes have vanishing horizon
radius, while here we used black holes whose horizon radius
remains finite in the extremal limit due to rotation. Note,
however, that the isometry group $SL(2,R)\times U(1)$ has a
natural superextension $SU(1,1|1)$.

%%%%%%%%%%%%%%%%%%%%%%%%%%%%%%%%%%%%%%%%%%%%%%%%%%%%%%%%%%%%%%%%%%%%%%%%%
\setcounter{equation}{0}
\section{Geodesics and test scalar field on RBR$^+$}

To probe the geometry (\ref{nhz}) of a maximally extended patch
let us consider  geodesics for a particle of mass $\mu$
(possibly zero). The Hamilton-Jacobi equation in the RBR 
spacetime fully separates, and the geodesic equations can
be integrated explicitly. The Killing vectors $L_1$  and
$L_\phi$ also apply to  $ RBR^+$   (\ref{nhz}) (in terms of
$t,\, \varphi$ ), generating the conserved energy and the
azimuthal component of the momentum \be {\cal E}=-{\dot{x}}^\nu
g_{\nu t},\quad L = {\dot{x}}^\nu g_{\nu \varphi}, \ee where
dots denote derivatives with respect to the affine parameter
$\lambda$ on the geodesics. The constraint equation
$g_{\mu\nu}\dot{x}^\mu \dot{x}^\nu=\varepsilon$ with
$\varepsilon=1$ for timelike and $\varepsilon=0$ for null
geodesics, reads \be (x^2+1)^{-1}\left({\dot{x}}^2-(\cE
+xL)^2\right) +L^2 \sin^{-2}\theta +
{\dot{\theta}}^2+\varepsilon =0. \ee This separates to the
$\theta$-equation \be\label{teta}
 {\dot{\theta}}^2+ L^2 \sin^{-2}\theta={\cal K}^2,
\ee where the separation constant ${\cal K}^2\geq L^2$ at the
right hand side has the meaning of the total angular momentum,
and the radial equation \be\label{iks} {\dot{x}}^2 + ({\cal
K}^2+\varepsilon)(1+x^2)-(\cE+xL)^2=0. \ee Integration of
(\ref{teta}) gives \be \cos\theta=\sqrt{1-L^2/{\cal K}^2}
\,\cos(K\lambda), \ee while integration of (\ref{iks}) \be
x=\frac{\cE L}{q^2}+a\cos q\lambda\,, \ee where \be q^2={\cal
K}^2-L^2+\varepsilon\,,\quad a^2q^4=({\cal
K}^2+\varepsilon)(\cE^2-q^2). \ee Geodesics remain confined,
unless $q=0$, which happens when \be \varepsilon=0\,,\quad
{\cal K}^2=L^2. \ee This corresponds to the equatorial
propagation ($\theta=\pi/2$) of a massless particle. Coming
back to Eq.~(\ref{iks}) we find in this case \be
\lambda=\int\frac{dx}{\sqrt{\cE^2-L^2+2x\cE L}}. \ee From this
equation it is clear that null geodesics can escape to infinity,
but only for infinite values of the affine parameter. So the
spacetime is geodesically complete.

Let us now investigate the modes of a minimally coupled scalar
field. The inverse metric for (\ref{nhz}) reads \be
g^{\nu\lambda}\p\nu\p\lambda= \frac1{1+x^2}\left(-\p t ^2+
\frac{1-x^2\cos^2\theta}{\sin^2\theta} \p{\varphi}^2 +2x\p t
\p\varphi\right)+(1+x^2)\p{x}^2+\p\theta^2, \ee while
$\sqrt{-g}=\sin\theta$, so that the Klein-Gordon equation for
the modes $\psi=\psi(x,\theta)e^{i(m\varphi-\omega t )}$ takes
the form
\[
 \frac1{\sqrt{-g}}\p\nu\left(\sqrt{-g}g^{\nu\lambda}\p\lambda\psi\right)=
 \p{x}\left[(1+x^2)\p{x}\psi\right]
 +\frac1{\sin\theta}\p\theta\left(\sin\theta\p\theta\psi\right)+
\]
\be \frac{(\omega+mx)^2}{1+x^2}\psi-
\frac{m^2}{\sin^2\theta}\psi=\mu^2\psi. \ee This equation
clearly separates \be \psi(x,\theta)=\Theta(\theta)X(x), \ee
splitting into the pair of ordinary differential equations \be
\lb{Zz} \p{x}\left[(1+x^2)\p{x}X\right]
+\frac{(\omega+mx)^2}{1+x^2}X-\mu^2 X={\cal K}^2 X, \ee \be
 -\frac1{\sin\theta}\p\theta\left(\sin\theta\p\theta\Theta\right)
+\frac{m^2}{\sin^2\theta}\Theta={\cal K}^2\Theta \ee (we use
the same symbol for the total angular momentum as for
geodesics). The angular equation is that for the associated
Legendre functions, so the separation constant is \be {\cal
K}^2=l(l+1),\quad l\geq |m|,\quad l=1,2,\ldots. \ee The radial
equation after the change of the variable \be
x=i\frac{\xi+1}{\xi-1} \ee reduces to the hypergeometric type
equation \be \xi^2(\xi-1)^2 \p\xi^2 X+\xi(\xi-1)^2\p\xi X
-\frac14 (\omega_+\xi-\omega_-)^2 X-({\cal K}^2+\mu^2)\xi X =0,
\ee where $\omega_\pm=\omega\pm im$. Its general solution reads
\be\lb{hyp} X=C_1\xi^\alpha (1-\xi)^h F(a_+, b_+; c_+;\xi)+
C_2\xi^{-\alpha} (1-\xi)^h F(a_-, b_-; c_-;\xi) \ee with \be
\alpha=\frac12(\omega-im),\quad h=\frac12(1+\lambda), \quad
a_\pm =h\pm\omega,\quad b_\pm=h\mp im,\quad c_\pm=1\pm2\alpha,
\ee and \be\lb{lam} \lambda=\sqrt{1+4({\cal K}^2+\mu^2-m^2)}.
\ee When $x\to \pm\infty$ the argument $\xi$ of the
hypergeometric functions goes to 1, and both diverge as
$(1-\xi)^{-\lambda}$ making the solution non-square integrable,
unless the series reduces to a polynomial for $a_{\pm} = h
\pm\omega=-n$ with $n$ integer. Choosing $\omega>0$ (in which
case $C_1=0$), this condition reads \be \label{spect}
\omega=n+\frac12 +\sqrt{\left(l+\frac12\right)^2-m^2 +\mu^2}.
\ee One sees that the energy spectrum of the minimally coupled
scalar field is discrete both for massive and massless
particles. Recall that in the latter case classical geodesics
can escape to infinity, but this takes an infinite time.  Note
that the spectrum is not degenerate with respect to the
azimuthal quantum number $m$, which could be expected for lack
of spherical symmetry. For $m=0$ we recover from our formula
(\ref{spect}) the usual result for quantization on $AdS_2\times
S^2$.

For the near horizon limit of the extremal Kerr-(Newman)
\cite{BaHo99} one finds essentially the same radial equation,
but a different angular equation. In that case the separation
constant ${\cal K}^2$ is such that for certain modes $\lambda$
is complex. This corresponds to the continuous spectrum. It was
shown in $\cite{BaHo99}$ that in this part of the spectrum
there are superradiant modes inherited from the Kerr field.

Coming back to the spectrum (\ref{spect}), one can see that it
is equidistant, which may indicate that it can be obtained
purely algebraically. Indeed, the anti-de Sitter group $SL(2,R)$
may be used as a dynamical symmetry. In other terms, one can
put the $x, \,t$  part of the wave equation into the form of a
conformal quantum mechanics.

%%%%%%%%%%%%%%%%%%%%%%%%%%%%%%%%%%%%%%%%%%%%%%%%%%%%%%%%%%%%%%%%%%%%%%%%%%%%%%%%%%%%%%%

\setcounter{equation}{0}
\section{Conformal mechanics}
Let us show that the $RBR^0$ geometry generates a modified
version of the relativistic conformal mechanics found recently
in \cite{Cl98} as a generalization of the de
Alfaro-Fubini-Furlan non-relativistic model. One starts by
multiplying the interval (\ref{bremda}) par $M^2=$ const and
rescaling time by $\tau \to M^{-1}\tau$ to arrive at the metric
\be ds^2=-r^2 d\tau^2 + \left(\frac{M}{r}\right)^2 dr^2
+M^2\left[d\theta^2+\sin^2\theta (d\phi+rd\tau/M)^2\right]. \ee
The associated Klein-Gordon operator is \be -p^2 \equiv
\frac1{\sqrt{-g}}p_{\nu}\sqrt{-g}g^{\nu\lambda}p_{\lambda} =
M^{-2}(-{\cal C} + {\cal K}^2 - L^2), \ee where \be
L=-p_\phi,\quad {\cal K}^2=p_\theta^2+\sin^{-2}\theta p_\phi^2,
\ee and \be {\cal C} = L_0^2 - \frac12(L_1L_{-1} + L_{-1}L_1)
\ee is the Casimir operator for the $sl(2,R)$ algebra.

Define the Hamiltonian as the time-component $p_{\tau}$ of the
four-momentum. The mass-shell constraint $p^2=\mu^2$ may be
solved to eliminate $p_{\tau}$ in terms of $r$, $p_r$ and the
angular quantum numbers $K$ and $L$. Introducing for
convenience a new variable \be \rho=\frac{2M}{r^{1/2}}, \ee so
that the RBR$^0$ metric takes the form \be
ds^2=-\left(\frac{2M}{\rho}\right)^4 d\tau^2 +
\left(\frac{2M}{\rho}\right)^2 d\rho^2
+M^2\left[d\theta^2+\sin^2\theta (d\phi+4M
d\tau/\rho^2)^2\right], \ee we thus obtain the Hamiltonian \be
H=\left(\frac{2M}{\rho}\right)^2\left\{\sqrt{\mu^2+(\rho^2
p_\rho^2+4{\cal K}^2)/(2M)^2} -L/M\right\}. \ee Following
\cite{Cl98} one can rewrite this as \be H=\frac{p_\rho^2}{2f} +
\frac{\mu g}{2f\rho^2} \ee with \be
f=\frac12\left(\sqrt{\mu^2+(\rho^2 p_\rho^2+4{\cal
K}^2)/(2M)^2}+L/M\right), \ee where \be g=4M^2\mu + 4({\cal
K}^2-L^2)/\mu. \ee This is precisely the  expression found in
\cite{Cl98} for a {\em charged} particle in the
Bertotti-Robinson spacetime and the corresponding
electromagnetic field. In our case the particle is uncharged,
but comparing with \cite{Cl98} one can see that now the
normalized azimuthal momentum $L/M$ enters $f$ in the same way
as the electric charge in \cite{Cl98}. This is the only
modification to the usual Bertotti-Robinson case which is
introduced by the RBR geometry. The DFF mechanics of \cite{DFF}
is recovered in the non-relativistic limit $M\to\infty,\,
L/M\to \mu$ keeping $g$ fixed. Then $f\to\mu$ and we get
\be\lb{nrDFF} H=\frac{p_{\rho}^2}{2\mu}+\frac{g}{2\rho^2}. \ee

From the relativistic conformal mechanics point of view, the
three generators of the $SL(2,R)$ symmetry are \be H =
iM^{-1}L'_1, \quad D = 2iL'_0, \quad K  = iML'_{-1}, \ee where
the reduced generators $L'_n$ are obtained from the Killing
vectors (\ref{kils}) by eliminating $p_{\tau} =
H(\rho,p_{\rho})$ and the conjugate variable $\tau =0$. The
generators of dilations $D$ and of special conformal
transformations $K $ are given by \be D=\rho p_\rho,\quad K
=\frac12 f \rho^2. \ee These are the spectrum generating
operators which, acting on the eigenstates of $H$, produce
other eigenstates. It is easy to show that the action of the
exponential operator $\exp(i\alpha D)$ on a state with an
energy $\omega$  gives an eigenstate with the energy
$\exp(2\alpha)\omega$, this means that the energy spectrum is
continuous. In addition, the lowest energy eigenstate is not
normalizable. This may seem strange in view of the result which
we obtained in the previous section for the Klein-Gordon
equation. Note, however, that we have used a different
coordinate patch, namely,RBR$^+$. The crucial difference is
that this patch does not have a horizon. The explanation  comes
in exactly the same way as for the ordinary Bertotti-Robinson
geometry \cite{GiTo99}.

First recall that already in \cite{DFF} it was argued that one
can redefine the Hamiltonian in such a way that the spectrum
becomes discrete. Any linear combination of the three generators
$H,\,D,\,K $ can be used as an evolution operator since all of
them generate transformations leaving the action invariant. The
particular combination \be \label{HK } H'=H+K  \ee was suggested
in \cite{DFF} as a `better' Hamiltonian with discrete spectrum.
Indeed, (\ref{HK }) amounts to adding  a quadratic potential to
$H$. Gibbons and Townsend \cite{GiTo99} noted that, in the
context of the BR  geometry, the transition from the
(relativistic version of) $H$ to $H'$ is equivalent to a
redefinition of the Hamiltonian with respect to the time
variable on the globally defined coordinate patch (see also
\cite{Vo99}). We shall now verify that this remains true for the
 RBR  geometry as well, thus explaining why we have obtained a
discrete spectrum for a test  scalar field on the background of
AdS$^+$, and show that the $sl(2,R)$ algebra expressed in
$t,\,x$ coordinates acts indeed as a spectrum generating
algebra.

We expect that the raising and lowering spectrum generating
operators ${\cal L}_{\pm 1}$ will be hermitean conjugate to each
other, and we wish to choose the hamiltonian\footnote{We use in
the following units such that $M = 1$.} as ${\cal H}={\cal
L}_0=i\p{t}$. Then the $sl(2,R)$ algebra can be recast into the
following form \ba
{\cal L}_0&=&\frac{i}{2}({ L}_1+{ L}_{-1}),\nonumber\\
{\cal L}_{\pm 1}&=&\frac{i}{2}({ L}_1-{ L}_{-1})\mp L_0, \ea so
that again \be\label{calsl} [{\cal L}_n,\,{\cal
L}_m]=(n-m){\cal L}_{n+m}, \ee but now (the $L_n$ being
antihermitean) ${\cal L}_0$ is hermitean ${\cal L}_0^+={\cal
L}_0$, while ${\cal L}_{1}^+={\cal L}_{- 1}$.

Let us introduce for AdS$^0$ the `light cone' coordinates
$t^\pm$: \be \tau=\frac12(t^++t^-),\quad r=2(t^+-t^-)^{-1}. \ee
Then passing from AdS$^0$ to AdS$^+$ amounts to the following
transformation \be t^\pm=\tan u^\pm, \ee and it can be checked
that our previous AdS$^+$ coordinates are related to these as
\be t=u^++u^-,\quad \arctan (x^{-1})=u^+-u^-. \ee For the
angular part we have \be \phi=\varphi +\ln\left|\frac{\cos
u^+}{\cos u^-}\right|, \ee so the entire metric reads \be
ds^2=-\frac{4du^+du^-}{\sin^2(u^+-u^-)}
+d\theta^2+\sin^2\theta\left(d\varphi+\cot(u^+ - u^-)(du^+ +
du^-) \right)^2. \ee In the new variables the above generators
take the more symmetric form \ba
{\cal L}_0&=&\frac{i}{2}(\p{+}+\p{-}),\nonumber\\
{\cal L}_{\pm 1}&=&\frac{i}{2}\left[\e^{\pm 2iu^{+}}(\p{+}\mp
i\p{\varphi}) +\e^{\pm 2iu^{-}}(\p{-}\pm i\p{\varphi})\right],
\ea with \be\lb{ppm} \p{\pm} \equiv \p{u^{\pm}} = \p{t} \mp
(1+x^2)\p{x}. \ee Restricting them to modes depending on
$\varphi$ as $\e^{im\varphi}$, one obtains \be {\cal L}_{\pm
1}^{(m)}=\frac{i}{2}\left[\e^{\pm 2iu^{+}}(\p{+}\pm m) +\e^{\pm
2iu^{-}}(\p{-}\mp m)\right]. \ee

To make contact with our previous treatment of the Klein-Gordon
equation let us write the $sl(2,R)$ Casimir operator: \ba &&
{\cal C} \equiv {\cal L}_0({\cal L}_0-1)-{\cal
L}_{-1}^{(m)}{\cal
L}_{1}^{(m)} =\nonumber\\
&&
-\sin^2(u^+-u^-)(\p{+}\p{-}+m^2)+\frac{im}{2}\sin[2(u^+-u^-)](\p{+}+\p{-}).
\ea For eigenmodes of ${\cal L}_0 = i\p{t}$ with eigenvalue
$\omega$, this is (according to (\ref{ppm})) identical to the
radial Klein-Gordon operator \be {\cal C} =
\p{x}(1+x^2)\p{x}+\frac{(\omega+mx)^2}{1+x^2} - m^2, \ee with
non-negative eigenvalues $(\lambda^2-1)/4$ (where $\lambda$ is
given in (\ref{lam})).

The representation of the spectrum generating algebra now can
be found in a standard way. Assuming (without loss of
generality) $\omega > 0$, the ground state of ${\cal L}_{0}$ is
annihilated by ${\cal L}_{1}$: \be {\cal
L}_{0}\phi_0=\omega_0\phi_0,\quad{\cal L}_{1}\phi_0 =0, \ee the
solution being \be \phi_0=\xi^{(im-h)/2}(1-\xi)^h,\quad
\xi=\e^{2i(u^+-u^-)},\quad \omega_0=h, \ee where
$h=(1+\lambda)/2$ is the conformal weight. As follows from the
commutation relations (\ref{calsl}), the operator ${\cal
L}_{-1}^{(m)}$ acts as a raising operator in the space of the
eigenvectors of $ {\cal L}_{0}$, shifting the eigenvalue by
unity, and consequently \be \omega_n=h+n. \ee Acting $n$ times
by \be {\cal L}_{-1}^{(m)} = (\xi^{-1/2}+\xi^{1/2}){\cal L}_0/2
- (\xi^{-1/2}-\xi^{1/2})(\xi\partial_{\xi} + im/2), \ee we find
the following expression for the $n-$th eigenfunction
\be\lb{hyp2} \phi_n =\xi^{(im-h-n)/2}(1-\xi)^h F(-n,
h+im;1-h-n+im;\xi), \ee where the hypergeometric function with
one of the first two indices equal to a negative integer is a
polynomial of degree $n$. This is exactly our previous solution
(\ref{hyp}) with $C_1 = 0$. In checking this formula it is
helpful to derive first the following operator identities \ba
{\cal L}_{-1}^{(m)}\xi^{-im/2} & = & \xi^{-im/2}{\cal
L}_{-1}^{(0)}, \nonumber\\
% &&\left[\e^{-2iu^+}\p{+}+\e^{-2iu^-}\p{-}\,,\,
% \left(\e^{2iu^-}-\e^{2iu^+}\right)^\nu\right] = 0,
\left[ \xi\partial_{\xi}\,,\,(\xi^{-1/2}-\xi^{1/2})^{-\omega}
\right] & = & \frac{\omega}2
\,\frac{(\xi^{-1/2}+\xi^{1/2})}{(\xi^{-1/2}-\xi^{1/2})}\,
(\xi^{-1/2}-\xi^{1/2})^{-\omega}, \ea
% the second one being valid for any $\nu$.
and to use the recurrence relation between hypergeometric
functions \be \frac{d}{d\xi}
\left(\xi^{c-1}(1-\xi)^{b-c+1}F(a,b;c;\xi)\right)=
(c-1)\xi^{c-2}(1-\xi)^{b-c}F(a-1,b;c-1;\xi), \ee arriving at
the recurrence between eigenfuctions \be {\cal L}_{-1}^{(m)}
\phi_n=(h+n-im)\phi_{n+1}. \ee

If one tries to put the $RBR^+$ problem into the form of a
relativistic  DFF conformal mechanics, the result can not be
expressed in algebraic form. For a discussion of this point
in the ordinary BR case see \cite{Mi01}.
%%%%%%%%%%%%%%%%%%%%%%%%%%%%%%%%%%%%%%%%%%%%%%%%%%%%%%%%%%%%%%%%%%%%%%%%%%%%%%%

\setcounter{equation}{0}
\section{ Asymptotic symmetry }
Now let us explore {\em asymptotic symmetries}, which are, by
definition, the coordinate transformations leaving the metric
invariant as $r\to\infty$. To find the corresponding generators
$\cal K$ one has to solve the Killing equations \be\lb{aki}
g_{\mu\nu,\lambda}{\cal K}^\lambda\;+\;g_{\mu\lambda}{\cal
K}^\lambda_{,\nu} \;+\;g_{\nu\lambda}{\cal K}^\lambda_{,\mu}=0
\ee assuming a suitable falloff for the metric near the
boundary. For BR these are given by the asymptotic symmetries
of $AdS_2$ together with the exact symmetry $SO(3)$ of the full
spacetime. The former were discussed in a number of papers
\cite{CaMi99,CaMi002, CaMi004,NaNa00,CaCa00}.  One assumes in
the $AdS_2$ sector (for $AdS^0$ patch) \be
g_{\tau\tau}=-r^2+O(1),\quad g_{rr}= 1/r^2 +O(1/r^4),\quad
g_{r\tau}=O(1/r^3). \ee The solution of (\ref{aki}) then reads
\be\lb{kBR} {\cal K}=\left(\epsilon(\tau)+\frac{1}{2
r^2}\epsilon''(\tau)\right)\p{\tau}- r\epsilon'(\tau)\p{r}, \ee
where $\epsilon(\tau)$ is an arbitrary (infinitesimal) real
function, and primes denote derivatives over $\tau$. These
transformations generate the Virasoro algebra. Indeed, expanding
$\epsilon(\tau)$ in the Laurent series \be
\epsilon(\tau)=\sum_{-\infty}^\infty \epsilon_n \tau^{(1-n)},
\ee we can write \be {\cal K}= \sum_{-\infty}^{\infty}
\epsilon_n L_n, \ee where $\epsilon_n$ are arbitrary real
numbers. The infinite set of generators \be\label{VirBR}
L_n=\tau^{-n}\left[\left(\tau+\frac{n(n-1)}{2r^2\tau}\right)\p{\tau}
\,+\,(n-1)r\p{r}\right], \ee satisfies the Virasoro algebra
without a central charge \be [L_n,L_m]=(n-m)L_{n+m} \ee up to
terms $O(r^{-4})$.

Now let us turn to the RBR spacetime, also choosing the $AdS^0$
patch. Actually we shall consider the more general class of
metrics 
\be\label{brkn}
 ds^2= F(\theta)\left(-r^2 d\tau^2+\frac{dr^2}{r^2}
+d\theta^2\right) + \frac{\sin^2\theta}{F(\theta)}(d\phi+\gamma r
d\tau)^2, 
\ee 
($\gamma$ constant), which includes the extreme
Kerr-Newman near-horizon metric \cite{BaHo99} as well as the
near-horizon metric for extreme rotating dilaton-axion black
holes with NUT charge \cite{ClGa01}. Irrespective of the factor
$F(\theta)$, the $SL(2,R)\times U(1)$ symmetry is enlarged to
an infinite-dimensional symmetry at the boundary $r\to\infty$
as well. We have to solve the Killing equations (\ref{aki})
assuming the following behavior for the metric as $r\to\infty$:
\ba 
&g_{\tau\tau} = - r^2(F-\gamma^2F^{-1}\sin^2\theta) + O(1),\qquad
g_{rr} = F/r^2 +O(1/r^4),&\nonumber \\
&g_{\phi\phi} = F^{-1}\sin^2\theta+O(1/r^2),\qquad
g_{\tau\phi} = \gamma r F^{-1}\sin^2\theta +O(1/r),&\nonumber \\
&g_{\theta\theta} = F + O(1/r^2),\quad  g_{r\tau}=O(1/r^3), \quad
g_{r\phi}=O(1/r^3),& 
\ea 
together with suitable conditions for
$g_{\theta\phi},\,g_{\theta r},\,g_{\theta\tau}$. The solution
reads 
\be\lb{kBREMDA} 
{\cal K}=\left(\epsilon(\tau)+\frac{1}{2
r^2}\epsilon''(\tau)\right)\p{\tau}- r\epsilon'(\tau)\p{r}+
\left(k - \gamma\frac{\epsilon''(\tau)}{r}\right)\p{\phi}, 
\ee
where $k$ is an arbitrary constant. The $k$-term decouples, it
generates an exact $U(1)$ symmetry of the bulk spacetime:
$L_\phi=\p{\phi}$. The remaining transformations form again the
Virasoro algebra. Carrying out a Laurent expansion one finds
\be\label{Vir}
L_n=\tau^{-n}\left[\left(t+\frac{n(n-1)}{2r^2\tau}\right)\p{\tau}
\,+\,r(n-1)\p{r}- \gamma\frac{n(n-1)}{r\tau}\p{\phi}\right]. 
\ee
Note the non-trivial mixing with $\p{\phi}$ due to the fact that
we no longer deal with the direct product of $AdS_2$ and a
sphere. The $sl(2,R)$ subalgebra of the Virasoro algebra
coincides with the exact symmetry of RBR exhibited in
(\ref{kils}). Further work is needed to find out whether the
representations of this algebra in terms of asymptotic
spacetime deformations acquire a central charge in the sense of
Brown and Henneaux \cite{BrHe86}.

%%%%%%%%%%%%%%%%%%%%%%%%%%%%%%%%%%%%%%%%%%%%%%%%%%%%%%%%%%%%%%%%%%%%%%%%%%%%%%%%%%%%%%%
\section{Concluding remarks}
Our results suggest that the set of geometries $AdS_n\times K$
with compact $K$ relevant for holography can be extended to
manifolds which do not have the structure of a direct product.
Several examples of such  geometries are provided as the near
horizon limits of various rotating black holes in four
dimensions. Such manifolds have the isometry $SL(2,R)\times
U(1)$ and they do not split into $AdS_2\times K$ even
asymptotically.

We have constructed explicitly  conformal mechanics associated
to  $RBR^0 $ as a modified relativistic De Alfaro-Fubini-Furlan
model in which the angular momentum plays the role of the
electric charge. We also checked that the transition to $RBR^+$,
similarly to the case of $AdS_2\times S^2$, redefines the time
variable in such a way that the new Hamiltonian has a discrete
spectrum bounded from below.  The associated $SL(2,R)$ symmetry
is naturally extended to the Witt algebra which is realized as
an asymptotic symmetry of the four-dimensional spacetime.

It is worth noting that although the conformal boundary is
singular in the rotating case, the asymptotic symmetry  still
contains the Virasoro algebra. So in this respect the only
difference with the case of the factorizable geometry
$AdS_2\times S^2$ whose boundary is non-singular is that the
action of the Virasoro generators cannot be separated from the
action of symmetries on $K$. This is valid both for our new
rotating BR spacetime, and for the previously considered
extremal Kerr(-Newman) throat. We intend to discuss the issue
of the central charge in a separate paper.

Our final remark concerns supersymmetry. Rotation breaks the
supersymmetry of $AdS_2\times S^2$, and our new solution is not
supersymmetric from the $N=4, d=4$ supergravity point of view.
However, the isometry group $AdS_2\times U(1)$ allows for a
superextension $SU(1,1|1)$. It remains to be clarified whether
supersymmetry still plays any role in our case.
\section*{Acknowledgements}
DG is grateful to LAPTH Annecy for hospitality and  to CNRS for
support which made possible this collaboration. His work was
also supported in part by the RFBR grant 00-02-16306.


\newpage
\begin{thebibliography}{20}

\bibitem{Ma} J.~M.~Maldacena,
Adv.~Theor.~Math.~Phys.~{\bf 2} (1998) 231; \\
S.~S.~Gubser, I.~R.~Klebanov and A.~M.~Polyakov,
Phys.~Lett.~{\bf B428} (1998) 105; \\
O.~Aharony, S.~S.~Gubser, J.~M.~Maldacena, H.~Ooguri and Y.~Oz,
Phys.~Rept.~{\bf 323} (2000) 183.
\bibitem{GiHu82}
G.~Gibbons and C.~M.~Hull, Phys. Lett.~{\bf B109} (1982) 190.
\bibitem{To83}
K.~P.~Tod, Phys. Lett.~{\bf B121} (1983) 241.
\bibitem{BoPeSk98}
H.~J.~Boonstra, B.~Peeters, and K.~Skenderis, Nucl. Phys. {\bf
B533} (1998) 127.
\bibitem{Yo99}
D. Youm, Phys.\ Rev.\ D {\bf 60} (1999) 064016; Nucl. Phys.
{\bf B573} (2000) 257.
\bibitem{CaCaCaMi01}
M.~Cadoni, P.~Carta, M.~Cavagli\`{a}, S.~Mignemi, {\sl
Conformal dynamics of 0-branes}, hep-th/0105113.
\bibitem{Cl98}
P. Claus, M. Derix, R. Kallosh, J. Kumar, P.K. Townsend and A.
Van Proeyen, Phys. Rev. Lett. {\bf 81} (1998) 4553.
\bibitem{Ka99} R.~Kallosh,
{\sl Black holes, branes and superconformal symmetry},
hep-th/9901095; {\sl Black holes and quantum mechanics},
hep-th/9902007.
\bibitem{DFF}
V. de Alfaro, S. Fubini and G. Furlan,
%{\sl Conformal invariance in quantum mechanics},
Nuovo Cim. {\bf 34A}, (1976) 569.
\bibitem{GiTo99}
G.~W.~Gibbons and P.~K.~Townsend,
% {\sl Black Holes and Calogero Models}
Phys.~Lett.~{\bf B454} (1999) 187.
\bibitem{MiSt99}
J.~Michelson and A.~Strominger, JHEP {\bf 9909} (1999) 005.
\bibitem{Az99}
J.A. de Azc{\'a}rraga, J.M. Izquierdo, J.C. P{\'e}rez-Bueno and
P.K. Townsend,
%{\sl Superconformal mechanics, black holes and non-linear realizations},
Phys.\ Rev.\ D {\bf 59} (1999) 084015.
\bibitem{AkKu99}
V. Akulov and M. Kudinov, Phys.~Lett.~{\bf B4260} (1999) 365.
\bibitem{Wy99}
N. Wyllard, J. Math. Phys. {\bf 41} (2000) 2826.
\bibitem{BeVa94}
E. Bergshoeff and M. Vasiliev, Int. Journ. Mod. Phys. {\bf A
10} (1995) 3477.
\bibitem{Ku99}
J.~Kumar, JHEP {\bf 9904} (1999) 006.
\bibitem{Ku00}
J. Kumar,
%{\sl Raiders of the lost AdS},
JHEP {\bf 0005} (2000) 035.
\bibitem{Ra01}
B. Rai, {\sl Dynamics of $AdS_2$ and enlargement of $SL(2,R)$ to
$c=1$ `cut-off' Virasoro algebra}, hep-th/0104142.
\bibitem{CaCaKl01}
M.~Cadoni, P.~Carta, and D.~Klemm,
% {\sl Large N limit of
%Calogero-Moser models and conformal field theories},
Phys.~Lett.~{\bf B503} (2001) 205.
\bibitem{MaSt97}
J.~Maldacena and A.~Strominger,
%{\sl Universal low-energy dynamics for rotating black holes},
Phys.\ Rev.\ D {\bf 56} (1997) 4975.
\bibitem{MaMiSt99}
J.~Maldacena, J.~Michelson, and A.~Strominger,
%{\sl Anti-de Sitter Fragmentation,}
JHEP {\bf 03} (1999) 011.
\bibitem{Mi98}
J.~Michelson, Phys.\ Rev.\ D {\bf 57} (1998) 1092.
\bibitem{Vo99}
R.~Britto-Pacumio, J.~Michelson, A.~Strominger  and A.~Volovich,
{\sl Lectures on superconformal quantum mechanics and
multi-black hole moduli space}, hep-th/9911066.
\bibitem{St99} A.~Strominger,
%$AdS_2$ {\sl Quantum Gravity and String Theory},
JHEP {\bf 01} (1999) 007.
\bibitem{CaMi99} M.~Cadoni and S.~Mignemi,
Phys.\ Rev.D {\bf 59} (1999) 081501; Nucl.~Phys.~{\bf B557}
(1999) 165.
\bibitem{NaNa00} J.~Navarro-Salas and P.~Navarro,
Nucl.~Phys.~{\bf B579} (2000) 250.
\bibitem{CaMi002} M.~Cadoni and S.~Mignemi,
%{\sl Symmetry breaking, central charges and the $AdS_2/CFT_1$
%correspondence},
Phys.~Lett.~{\bf B490} (2000) 131.
\bibitem{Ca99}
M.~Cadoni, Phys. Rev. D {\bf 60} (1999) 084016.
\bibitem{CaMi004} M.~Cadoni. P.~Carta and S.~Mignemi,
%{\sl A realization of the infinite-dimensional symmetries of
%conformal mechanics},
Phys.\ Rev.\ D {\bf 62} (2000) 086002.
\bibitem{CaCa00} M.~Cadoni and M.~Cavagli\`{a},
Phys. Lett. {\bf B499} (2001) 315; Phys. Rev. D {\bf 63} (2001) 084024.
\bibitem{CaCaKlMi00}
M.~Cadoni, P.~Carta, D.~Klemm, S.~Mignemi, Phys. Rev. D {\bf 63}
(2001) 125021.
\bibitem{CaCaVa00}
M.~Caldarelli, G.~Catelani and L.~Vanzo,
%{\sl Action, hamiltonian and CFT for 2D black holes},
JHEP {\bf 0010} (2000) 005.
\bibitem{CaVa00}
G.~Catelani and L.~Vanzo, {\sl On the $\sqrt{2}$ puzzle in
$AdS_2/CFT_1$}, hep-th/0009186.
\bibitem{CaCar01} M.~Cadoni and P.~Carta,
%{\sl The AdS/CFT correspondence in two dimensions},
Mod. Phys. Lett. {\bf A16} (2001) 171.
\bibitem{BrHe86}
J.~D.~Brown and M.~Henneaux, Comm.~Math~Phys. {\bf 104} (1986)
207.
\bibitem {Te01}
H.~Terashima, Phys. Rev. D {\bf 64} (2001) 064016.
\bibitem{St98} A.~Strominger,
%{\sl Black Hole Entropy from Near-Horizon Microstates},
JHEP {\bf 02} (1998) 009.
\bibitem{BaHo99} J. Bardeen and G.T. Horowitz,
%{\sl The extreme Kerr throat geometry: a vacuum analogue of $AdS_2 \times S^2$},
Phys. Rev. D {\bf 60} (1999) 104030.
\bibitem{Za98} O.~B.~Zaslavsky,
%{\sl Horizon/Matter Systems Near the Extreme State},
Class. Quant. Grav. {\bf15} (1998) 3251.
\bibitem{CvLa99} M.~Cveti\v{c} and F.~Larsen,  Nucl.\ Phys. 
{\bf B531} (1998) 239;
%{\sl Microstates of Four-Dimensional Rotating
%Black Holes from Near-Horizon Geometry},
Phys.\ Rev.\ Lett. {\bf 82} (1998) 484.
\bibitem{Be00}
D.~Berman, {\sl Aspects of holography and rotating AdS black
holes}, hep-th/0002235.
\bibitem{BePa00}
D.~S.~Berman and M.~K.~Parikh, Phys. Lett. {\bf B463} (1999)
168.
\bibitem{HoHuTa99}
S.~W.~Hawking, C.~J.~Hunter, and M.~M.~Taylor-Robinson, Phys.
Rev. {\bf D59} (1999) 064005.
\bibitem{CaKlSa01}
M.~M.~Caldarelli, D.~Klemm, and W.~Sabra, JHEP {\bf 0105} (2001) 014.
\bibitem{ClGa01} G\'erard Cl\'ement and Dmitri Gal'tsov, Phys. Rev. D
{\bf 63} (2001) 124011.
\bibitem{GaKe94}
D.~V.~Gal'tsov and O.~V.~Kechkin, Phys. Rev. D {\bf 50} (1994)
7394; {\sl Hidden symmetries in dilaton--axion gravity''}, in
``Geometry and Integrable models'', P.N. Pyatov and S.N.
Solodukhin (Eds.), World Scientific, Singapore 1996.
\bibitem{Mi01}
S. Mignemi, {\sl Black holes and conformal mechanics},
hep-th/0104175.
\end{thebibliography}
\end{document}